\documentstyle[12pt]{article}
\begin{document}
\newcommand{\Dslash}{\not \!\! D}
\newcommand{\pslash}{\not \!\! p}
\newcommand{\kslash}{\not \!\! k}
\newcommand{\qslash}{\not \!\! q}

\makeatletter
\@addtoreset{equation}{section}
\def\theequation{\thesection.\arabic{equation}}
\makeatother

\begin{flushright}{UT-951}
\end{flushright}
\vskip 0.5 truecm

\begin{center}
{\large{\bf Spin-Statistics Theorem in Path Integral Formulation}}
\end{center}
\vskip .5 truecm
\centerline{\bf Kazuo Fujikawa}
\vskip .4 truecm
\centerline {\it Department of Physics,University of Tokyo}
\centerline {\it Bunkyo-ku,Tokyo 113,Japan}
\vskip 0.5 truecm

\begin{abstract}
We present a coherent proof of the 
spin-statistics theorem in path integral formulation. 
The local path integral measure and Lorentz invariant local
Lagrangian, when combined with Green's functions defined in terms 
of time ordered products, ensure causality regardless of 
statistics. 
The Feynman's $m-i\epsilon$ prescription ensures the positive
energy condition regardless of statistics, and the 
abnormal spin-statistics relation 
for both of spin-0 scalar particles and spin-1/2 Dirac particles
is excluded if one imposes the positive norm condition in 
conjunction with Schwinger's action principle. The minus 
commutation relation between one Bose and one Fermi field arises 
naturally in path integral. The Feynman's 
$m-i\epsilon$ prescription also ensures a smooth continuation to
Euclidean theory, for which the use of the Weyl anomaly
 is illustrated to exclude the abnormal statistics for the scalar
and Dirac particles 
not only in 4-dimensional theory but also in 2-dimensional 
theory.     
\end{abstract}
\large

\section{Introduction}
The spin-statistics theorem is one of the basic theorems
in theoretical physics. It has a long history\cite{pauli36}-
\cite{schwinger}, and 
Pauli established the theorem in its standard form\cite{pauli}. 
In a Lorentz invariant local field theory, the theorem holds 
provided the following 3 conditions are satisfied:\\
1. The vacuum is the lowest energy state.\\
2. Field variables either commute or anti-commute at space-like 
separation.\\
3. Norm in the Hilbert space is positive definite.\\

A further 
refinement of the theorem, in particular, the logical 
independence of the spin-statistics theorem and CPT theorem, 
another fundamental theorem in local field theory, has been
shown  by L\"{u}ders and Zumino\cite{luders} and 
Burgoyne\cite{burgoyne}. The formulation of 
Wightman\cite{wightman} plays an
essential role here. All these approaches are based on the 
operator formalism. A comprehensive account of 
the spin-statistics theorem has been given recently in 
Ref.\cite{duck}.

Another formulation of quantum theory, namely, the Feynman
path integral (and Schwinger's action principle) is widely
used in the applications of modern field theory.  It is 
 desirable to show the spin-statistics theorem in 
path integral approach not only for its logical completeness but 
also for a pedagogical purpose.
Also, the particles with abnormal statistics, such as 
the Faddeev-Popov ghost\cite{faddeev} and 
the bosonic Dirac particle as the Pauli-Villars regulator (see, 
for example,\cite{fujikawa2}), are commonly used in path 
integral, but these particles do not violate the causality as
one might naively expect. 
The aim of the present note is to show that we can now 
 give a coherent proof of the
spin-statistics theorem in modern path integral formulation which 
incorporates the Grassmann numbers as an integral 
part\cite{berezin}.
 A salient feature of 
path integral formulation is that all the Green's functions 
are defined in terms of time ordered products, and the notion 
such as the Wightman function\cite{wightman} is not available, 
at least in a natural way.  

From a technical view point, 
the major difference of various approaches to the spin-statistics 
theorem lies in how to incorporate the positive energy condition 
(Condition 1 above). Pauli in his original paper\cite{pauli} 
used essentially an explicit form of energy-momentum tensor.
L\"{u}ders and Zumino\cite{luders} and also 
Burgoyne\cite{burgoyne} greatly simplified the analysis by simply 
imposing the positive energy condition, which in turn leads to
an analyticity of the Wightman function. The formulation of 
Wightman\cite{wightman} enjoys a mathematical rigor and 
generality, but it is not quite 
accessible to everybody who is interested in the applications 
of field theory. Besides, an explicit construction of non-trivial
 models remains as a difficult issue in Wightman's formulation.
We here emphasize the familiar Feynman's $m-i\epsilon$ 
prescription 
as a manifestation of positive energy condition, which simplifies 
the analysis. The Feynman's $m-i\epsilon$ prescription is here 
assigned  a more 
fundamental meaning than just representing  a specific boundary 
condition to reproduce the result of operator formalism: In any 
fixed time slice of 4-dimensional space-time,
positive energy particles propagate in forward time direction and
negative energy particles propagate in backward time direction. 
By this way, the positive energy condition up to any finite order 
 in perturbation theory is ensured 
regardless of statistics.

The plan of this note is as follows: We first briefly summarize 
 the basic requirement of the path 
integral measure. The ordinary complex numbers and the Grassmann
numbers satisfy the basic requirement to define a local path 
integral measure. In the classical level, the field variables are
thus either totally commuting or totally anti-commuting. 
The complex (including real) numbers naturally
give rise to Bose-Einstein statistics and the Grassmann
numbers give rise to Fermi-Dirac statistics after performing 
path integral. The minus commutation
relation between one Bose and one Fermi field arises naturally  
in path integral, since the complex numbers and Grassmann 
numbers commute in the classical level. 
The local path integral measure and Lorentz
invariant local Lagrangian together with propagators, which are 
always defined in terms of time ordered products, are shown to 
ensure  the causality (Condition 2 above) regardless of 
statistics.
If one employs Feynman's $m-i\epsilon$ prescription, which allows
a smooth continuation from Euclidean theory, the positive energy 
condition is ensured regardless of statistics.
The basic criterion to exclude abnormal spin-statistics
relation is thus the positive norm in the Hilbert space or 
positive probability  for scattering processes 
(Condition 3 above).

We thus examine
what happens if we apply the Grassmann numbers to spin-0 
particles, or the complex numbers to spin-1/2 particles in
path integral. In conjunction with Schwinger's 
action principle, it is shown that the indefinite metric appears 
for spin $0$ particles if one uses Grassmann
numbers, and the negative metric for the negative energy 
states appears if one uses  complex numbers for Dirac particles. 
The Feynman's $m-i\epsilon$ prescription also allows a smooth
continuation to Euclidean theory, for which we illustrate the 
use of the Jacobian factor\cite{fujikawa} related to the Weyl
 transformation, which is sensitive to statistics, to exclude 
the abnormal statistics for the scalar and Dirac particles 
not only in 4-dimensional theory but also in 2-dimensional 
theory.     
The natural reasoning of our path integral analysis 
corresponds to that of Feynman\cite{feynman} and 
Pauli\cite{pauli2}. This is in 
contrast to the analyses on the basis of operator formalism in 
the standard textbooks\cite{bjorken2},
which emphasize the acausal behavior for abnormal statistics.
See also Ref.\cite{peskin} for a variation of the argument.  

\section{Schwinger's action principle and path integral measure}

The basic requirement of the path integral measure is that 
it is ``translation'' invariant in the 
functional space. By considering an infinitesimal 
quantity $\epsilon$ in the conventional integral, we have
\begin{eqnarray}
\int_{-\infty}^{\infty} dx f(x)
&=&\int_{-\infty}^{\infty} d(x+\epsilon) f(x+\epsilon)
=\int_{-\infty}^{\infty} dx f(x+\epsilon)\nonumber\\
&=&\int_{-\infty}^{\infty} dx f(x)+
\epsilon\int_{-\infty}^{\infty} dx \frac{d}{dx}f(x)
\end{eqnarray}
namely,
\begin{equation}
\int_{-\infty}^{\infty} dx \frac{d}{dx}f(x)=0
\end{equation}
which states the vanishing integral of a derivative. 
The relation which appeared here
\begin{equation}
d(x+\epsilon)=dx
\end{equation}
becomes  the basic relation in path integral also.

To derive the basic requirement for the path integral measure,
we start with a 
path integral for a real scalar particle 
\begin{equation}
\langle 0|0\rangle_{J}=\int{\cal D}\phi\exp\{i\int d^{4}x
[\frac{1}{2}
\partial_{\mu}\phi(x)
\partial^{\mu}\phi(x)
-\frac{1}{2}m^{2}\phi(x)^{2}+\phi(x)J(x)]\}
\end{equation}
where we added the source term.
For the moment, we deal with an abstract definiton of the path 
integral measure ${\cal D}\phi$ without precise specification.

We then have the basic condition on the path integral
\begin{eqnarray}
&&\langle 0|\partial^{2}_{\mu}\hat{\phi}(x)
+m^{2}\hat{\phi}(x)-J(x)|0\rangle_{J}\nonumber\\
&&= \int{\cal D}\phi\{
\partial^{2}_{\mu} \phi(x)
+m^{2}\phi(x)-J(x)
\}\nonumber\\
&&\times\exp\{i\int d^{4}x[\frac{1}{2}\partial_{\mu}
\phi(x)
\partial^{\mu}\phi(x)
-\frac{1}{2}m^{2}\phi(x)^{2}+\phi(x)J(x)]\}\nonumber\\
&&=i\int{\cal D}\phi\frac{\delta}{\delta\phi(x)}
\exp\{i\int d^{4}x[\frac{1}{2}\partial_{\mu}\phi(x)
\partial^{\mu}\phi(x)
-\frac{1}{2}m^{2}\phi(x)^{2}+\phi(x)J(x)]\}\nonumber\\
&&=0
\end{eqnarray}
where the first equality in this relation is a result of
Schwinger's action principle
\begin{equation}
\frac{\delta}{\delta J(x)}\langle 0|0\rangle_{J}
=i\langle 0|\hat{\phi}(x)|0\rangle_{J},
\end{equation}
and 
\begin{equation}
\langle 0|\partial^{2}_{\mu}\hat{\phi}(x)
+m^{2}\hat{\phi}(x)-J(x)|0\rangle_{J}=0
\end{equation}
is the equation of motion in the language of operator 
formalism.

This basic relation is satisfied by the ``translational''
 invariance of the path integral measure in functional space
\begin{equation}
{\cal D}(\phi+\epsilon)={\cal D}\phi.
\end{equation}
Here, $\epsilon(x)$ is an infinitesimal arbitrary function.
This fact is understood by defining $\phi^{\prime}(x)
=\phi+\epsilon$ as follows:
\begin{eqnarray}
&&\int{\cal D}\phi
\exp\{i\int d^{4}x[\frac{1}{2}\partial_{\mu}\phi(x)
\partial^{\mu}\phi(x)
-\frac{1}{2}m^{2}\phi(x)^{2}+\phi(x)J(x)]\}
\nonumber\\
&&=\int{\cal D}\phi^{\prime}
\exp\{i\int d^{4}x[\frac{1}{2}\partial_{\mu}\phi^{\prime}(x)
\partial^{\mu}\phi^{\prime}(x)
-\frac{1}{2}m^{2}\phi^{\prime}(x)^{2}+\phi^{\prime}(x)J(x)]\}
\nonumber\\
&&=\int{\cal D}\phi
\exp\{i\int d^{4}x[\frac{1}{2}\partial_{\mu}\phi(x)
\partial^{\mu}\phi(x)
-\frac{1}{2}m^{2}\phi(x)^{2}+\phi(x)J(x)]\nonumber\\
&&-i\int d^{4}x\epsilon(x)[\partial^{2}_{\mu}\phi(x)+m^{2}\phi(x)
-J(x)] \}
\end{eqnarray}
where the first equality is the statement that the naming of
integration variables does not change the integral itself.
This relation gives in the order linear in $\epsilon(x)$ 
\begin{eqnarray}
&&\int{\cal D}\phi i\int d^{4}y\epsilon(y)
[\partial^{2}_{\mu}\phi(y)+m^{2}\phi(y)
-J(y)]\nonumber\\
&&\times\exp\{i\int d^{4}x[\frac{1}{2}\partial_{\mu}\phi(x)
\partial^{\mu}\phi(x)
-\frac{1}{2}m^{2}\phi(x)^{2}+\phi(x)J(x)]\}=0.
\end{eqnarray}
If one chooses $\epsilon(y)$ as a $\delta$-functional one with 
a peak at $x$, one satisfies the basic requirement of the 
action principle (2.5). This analysis is valid for interacting
fields also.

We thus learn that the ``translation'' invariance of the path 
integral measure is equivalent to the equation of 
motion in operator formalism.
At this moment, two viable definitions of the path integral 
measure are known:\\
1. The first is a generalization of the ordinary integral,
which is ``translation'' invariant, by 
using the real or complex field variables
\begin{equation}
{\cal D}\phi\equiv \prod_{x}d\phi(x)
\end{equation}
namely we integrate over the field variable $d\phi(x)$,
which is an ordinary number, at each space-time point.
The ordinary complex or real numbers satisfy 
\begin{equation}
[\phi(x),\phi(y)]=0
\end{equation}
which leads to Bose-Einstein statistics after performing path 
integral.
\\
2. The second choice is to regard the variable $\psi(x)$ as 
Grassmann numbers defined at each point of space-time.(We 
here use $\psi(x)$ for Grassmann variables,
just for notational convenience.)
The integral is then defined as the (left-)derivative with 
respect to the Grassmann numbers
\begin{equation}
{\cal D}\psi\equiv \prod_{x}\frac{\delta}{\delta\psi(x)}
\end{equation}
which is also ``translation'' invariant in the functional 
space.
The Grassmann numbers anti-commute with themselves
\begin{equation}
\{\psi(x),\psi(y)\}=\psi(x)\psi(y)+\psi(y)\psi(x)=0
\end{equation}
and thus give rise to Fermi-Dirac statistics after 
performing path integral.

Field variables are either totally commuting or totally
anti-commuting in the classical level, and the propagators 
we use in path integral are  defined 
in terms of time ordered product. The appearance of the 
time ordered product is most easily understood if one formulates 
path integral starting with the evolution operator
\begin{equation}
\langle f|\exp[-i\hat{H}(t_{f}-t_{i})]|i\rangle
\end{equation}
and time slicing. It is important to recognize that our 
path integral measure (2.11) or (2.13) is {\em local} in the 
sense that path integral variables at each 
space-time point are allowed to change independently.
The space-time correlation of field variables after path integral 
is thus what the Lorentz 
invariant local Lagrangian implies. It is shown later that the
 time ordered product combined with the local path integral 
measure and Lorentz 
invariant local Lagrangian ensure the basic requirement of 
causality (Condition 2 above) regardless of statistics. 

We also emphasize  that Grassmann numbers and ordinary
complex numbers commute at the classical level, which naturally 
leads to commuting quantized variables after performing path 
integral. This property is not obvious in the operator
formulation\cite{luders}.

\section{Fermi-Dirac statistics for spin 0}

We now illustrate the difficulty if one uses Grassmann numbers 
for spin-0 real particles in path integral. The complex 
scalar field is written in terms of real scalar fields, and  
we first examine a real scalar field
\begin{equation}
\phi(x)^{\dagger}=\phi(x).
\end{equation}
If one writes the ordinary classical Lagrangian in terms of 
Grassmann numbers, one obtains
\begin{equation}
{\cal L}=\frac{1}{2}\partial_{\mu}\phi(x)
\partial^{\mu}\phi(x)
-\frac{1}{2}m^{2}\phi(x)^{2}=0
\end{equation}
namely, one cannot define a meaningful theory. This is a path
integral version of the 
standard argument against the fermionic interpretation of
spin $0$ particles in \cite{pauli36}-\cite{pauli} and 
\cite{luders}-\cite{duck}.

To avoid this difficulty, one may use a pair of real Grassmann
fields
$\xi(x)$ and $\eta(x)$
\begin{equation}
{\cal L}=\partial_{\mu}\xi(x)
\partial^{\mu}\eta(x)
-m^{2}\xi(x)\eta(x)
\end{equation}
where
\begin{equation}
\xi(x)^{\dagger}=\xi(x), \ \ \ 
\eta(x)^{\dagger}=\eta(x).
\end{equation}
The Faddeev-Popov ghost fields in gauge theory have this structure
\cite{faddeev}. (One may equally consider a complex scalar
defined by $\varphi(x)=(\xi(x)+i\eta(x))/\sqrt{2}$).
To ensure the unitarity of S-matrix,
$SS^{\dagger}=S^{\dagger}S=1$, the Lagrangian need to
be hermitian. The above Lagrangian gives
\begin{eqnarray}
{\cal L}^{\dagger}&=&\partial^{\mu}\eta(x)
\partial_{\mu}\xi(x)
-m^{2}\eta(x)\xi(x)\nonumber\\
&=&-\partial^{\mu}\xi(x)
\partial_{\mu}\eta(x)
+m^{2}\xi(x)\eta(x)\nonumber\\
&=&-{\cal L}
\end{eqnarray}
and thus we have to add an extra imaginary factor $i$
\begin{equation}
{\cal L}=i\partial_{\mu}\xi(x)
\partial^{\mu}\eta(x)
-im^{2}\xi(x)\eta(x).
\end{equation}
Path integral in this case is defined by
\begin{equation}
\int{\cal D}\xi{\cal D}\eta\exp\{i\int d^{4}x
[i\partial_{\mu}\xi(x)
\partial^{\mu}\eta(x)
-im^{2}\xi(x)\eta(x)]\}.
\end{equation}
To derive the propagator, we add sources which are Grassmann
numbers
\begin{equation}
S_{J}=\int d^{4}x\{i\partial_{\mu}\xi(x)\partial^{\mu}\eta(x)
-im^{2}\xi(x)\eta(x)+\xi(x)J_{\xi}+J_{\eta}\eta(x) \}
\end{equation}
and consider the change of variables
\begin{eqnarray}
&&\xi(x)=\xi^{\prime}(x)
-J_{\eta}\frac{i}{\partial_{\mu}\partial^{\mu}+m^{2}-i\epsilon},
\nonumber\\
&&\eta(x)=\eta^{\prime}(x)
-\frac{i}{\partial_{\mu}\partial^{\mu}+m^{2}-i\epsilon}J_{\xi}.
\end{eqnarray}
The action is then written as 
\begin{equation}
S_{J}=\int d^{4}x\{i\partial_{\mu}\xi^{\prime}(x)
\partial^{\mu}\eta^{\prime}(x)
-im^{2}\xi^{\prime}(x)\eta^{\prime}(x)
-J_{\eta}\frac{i}{\partial_{\mu}\partial^{\mu}+m^{2}-i\epsilon}
J_{\xi} \}
\end{equation}
and the path integral is written as
\begin{eqnarray}
Z(J)&=&\int{\cal D}\xi{\cal D}\eta
\exp\{i\int d^{4}x[i\partial_{\mu}\xi^{\prime}(x)
\partial^{\mu}\eta^{\prime}(x)
-im^{2}\xi^{\prime}(x)\eta^{\prime}(x)\nonumber\\
&&-J_{\eta}\frac{i}{\partial_{\mu}\partial^{\mu}+m^{2}-i\epsilon}
J_{\xi}]\}\nonumber\\
&=&\int{\cal D}\xi^{\prime}{\cal D}\eta^{\prime}
\exp\{i\int d^{4}x[i\partial_{\mu}\xi^{\prime}(x)
\partial^{\mu}\eta^{\prime}(x)
-im^{2}\xi^{\prime}(x)\eta^{\prime}(x)\nonumber\\
&&-J_{\eta}\frac{i}{\partial_{\mu}\partial^{\mu}+m^{2}-i\epsilon}
J_{\xi}]\}
\end{eqnarray}
by using the translational invariance of the measure
${\cal D}\xi{\cal D}\eta
={\cal D}\xi^{\prime}{\cal D}\eta^{\prime}$.
The propagator is then given by
\begin{eqnarray}
&&\langle0|T^{\star}\hat{\xi}(x)\hat{\eta}(y)|0\rangle\nonumber\\
&&=
\frac{1}{Z}\int{\cal D}\xi{\cal D}\eta
\xi(x)\eta(y)\exp\{i\int d^{4}x[i\partial_{\mu}\xi(x)
\partial^{\mu}\eta(x)-im^{2}\xi(x)\eta(x)]\}\nonumber\\
&&=
\frac{-1}{i}\frac{\delta}{\delta J_{\xi}(x)}\frac{1}{i}
\frac{\delta}{\delta J_{\eta}(y)}
\ln Z(J)|_{J=0}\nonumber\\
&&=(-i)
\frac{i}{\partial_{\mu}\partial^{\mu}+m^{2}-i\epsilon}
\delta^{4}(x-y).
\end{eqnarray}
where the operator expression in the left-hand side is a
result of Schwinger's action principle. This propagator 
reflects precisely the local structure of the Lagrangian, since
the path integral measure is local.
Note that we use the same Feynman's 
$m^{2}-i\epsilon$ prescription as ordinary particles, which 
ensures a smooth continuation from Euclidean theory. This 
$m^{2}-i\epsilon$ prescription
dictates the propagation of the negative energy solution in 
the negative time direction and ensures the positive energy 
condition (Condition 1). 

One can first confirm the causality on
the basis of Bjorken-Johnson-Low (BJL) prescription.
By using the relation
$\delta^{4}(x-y)=\int\frac{d^{4}k}{(2\pi)^{4}}\exp[-ik(x-y)]$,
one can write the above propagator as  
\begin{eqnarray}
&&\langle0|T^{\star}\hat{\xi}(x)\hat{\eta}(y)|0\rangle\nonumber\\
&&=(i)
\int\frac{d^{4}k}{(2\pi)^{4}}\exp[-ik(x-y)]
\frac{i}{k_{\mu}k^{\mu}-m^{2}+i\epsilon}
\end{eqnarray}
or equivalently
\begin{equation}
\int d^{4}x\exp[ik(x-y)]\langle0|T^{\star}\hat{\xi}(x)
\hat{\eta}(y)|0\rangle
=\frac{-1}{k_{\mu}k^{\mu}-m^{2}+i\epsilon}.
\end{equation}
We now employ the BJL definition of $T$-product\cite{bjorken}: We
 can replace 
$T^{\star}$-product by the conventional $T$-product if the 
following condition is satisfied 
\begin{equation}
\lim_{k^{0}\rightarrow\infty}
\int d^{4}x\exp[ik(x-y)]\langle0|T^{\star}\hat{\xi}(x)
\hat{\eta}(y)|0\rangle=0,
\end{equation}
namely, the equal time limit (i.e., 
$k^{0}\rightarrow\infty$ limit) is well defined for the 
$T$-product. If the limit (3.15) does not vanish, the $T$-product
is defined by subtracting it from $T^{\star}$-product.
This gives a general definition of $T$ product.

The above propagator (3.14) satisfies this condition, and we have
\begin{eqnarray}
&&(-ik^{0})
\int d^{4}x\exp[ik(x-y)]\langle0|T\hat{\xi}(x)
\hat{\eta}(y)|0\rangle\nonumber\\
&&=
\int d^{4}x\exp[ik(x-y)]\nonumber\\
&&\times\frac{\partial}{\partial x^{0}}
[\theta(x^{0}-y^{0})\langle0|\hat{\xi}(x)\hat{\eta}(y)|0\rangle
-\theta(y^{0}-x^{0})\langle0|\hat{\eta}(y)\hat{\xi}(x)|0\rangle]
\nonumber\\
&&=\int d^{4}x\exp[ik(x-y)]\delta(x^{0}-y^{0})
\langle0|\{\hat{\xi}(x), \hat{\eta}(y)\}|0\rangle\nonumber\\
&&+
\int d^{4}x\exp[ik(x-y)]\langle0|T\partial_{x^{0}}
\hat{\xi}(x)\hat{\eta}(y)|0\rangle\nonumber\\
&&=\frac{ik^{0}}{k_{\mu}k^{\mu}-m^{2}+i\epsilon}
\end{eqnarray}
where we used the fact that Grassmann numbers are anti-commuting
$\xi(x)\eta(y)=-\eta(y)\xi(x)$.
By taking the limit $k^{0}\rightarrow\infty$ in this expression
and remembering the definition of $T$-product, we conclude
\begin{eqnarray}
&&\int d^{4}x\exp[ik(x-y)]\delta(x^{0}-y^{0})
\langle0|\{\hat{\xi}(x), \hat{\eta}(y)\}|0\rangle=0,\nonumber\\
&&\int d^{4}x\exp[ik(x-y)]\langle0|T
\partial_{x^{0}}\hat{\xi}(x)\hat{\eta}(y)|0\rangle
=\frac{ik^{0}}{k_{\mu}k^{\mu}-m^{2}+i\epsilon}.
\end{eqnarray}
Repeating the same procedure for the second expression in 
(3.17), we have
\begin{eqnarray}
&&(-ik^{0})
\int d^{4}x\exp[ik(x-y)]\langle0|T\partial_{x^{0}}\hat{\xi}(x)
\hat{\eta}(y)|0\rangle\nonumber\\
&&=\int d^{4}x\exp[ik(x-y)]\delta(x^{0}-y^{0})
\langle0|\{\partial_{x^{0}}\hat{\xi}(x), \hat{\eta}(y)\}|0\rangle
\nonumber\\
&&+\int d^{4}x\exp[ik(x-y)]\langle0|T
\partial^{2}_{x^{0}}\hat{\xi}(x)\hat{\eta}(y)|0\rangle
\nonumber\\
&&=\frac{(k^{0})^{2}}{k_{\mu}k^{\mu}-m^{2}+i\epsilon}
\end{eqnarray}
and considering the limit $k^{0}\rightarrow\infty$, we conclude
\begin{eqnarray}
&&\int d^{4}x\exp[ik(x-y)]\delta(x^{0}-y^{0})
\langle0|\{\partial_{x^{0}}\hat{\xi}(x), \hat{\eta}(y)\}|0\rangle
=1,\nonumber\\
&&\int d^{4}x\exp[ik(x-y)]\langle0|T
\partial^{2}_{x^{0}}\hat{\xi}(x)\hat{\eta}(y)|0\rangle
=\frac{\vec{k}^{2}+m^{2}}{k_{\mu}k^{\mu}-m^{2}+i\epsilon}.
\end{eqnarray}
Using the operator equation
of motion $(\partial_{\mu}\partial^{\mu}+m^{2})\hat{\xi}(x)=0$,
the last equation in (3.19) is written as
\begin{eqnarray}
&&\int d^{4}x\exp[ik(x-y)]\langle0|T
(\partial_{l}\partial^{l}-m^{2})\hat{\xi}(x)\hat{\eta}(y)|0\rangle
\nonumber\\
&&=-(\vec{k}^{2}+m^{2})\int d^{4}x\exp[ik(x-y)]\langle0|T
\hat{\xi}(x)\hat{\eta}(y)|0\rangle\nonumber\\
&&=-(\vec{k}^{2}+m^{2})\frac{-1}{k_{\mu}k^{\mu}-m^{2}+i\epsilon}
\end{eqnarray}
and  coming full circle back to the original equation 
(3.14).
We thus obtain the anti-commutation relations
\begin{eqnarray}
&&\delta(x^{0}-y^{0})
\langle0|\{\hat{\xi}(x), \hat{\eta}(y)\}|0\rangle=0,\nonumber\\
&&\delta(x^{0}-y^{0})
\langle0|\{\partial_{x^{0}}\hat{\xi}(x), \hat{\eta}(y)\}|0\rangle
=\delta^{4}(x-y)
\end{eqnarray}
and the causality (Condition 2) is ensured.

We here note that an  $i\epsilon$ prescription
different from (3.14) such as 
\begin{equation}
\int d^{4}x\exp[ik(x-y)]\langle0|T^{\star}\hat{\xi}(x)
\hat{\eta}(y)|0\rangle
=\frac{-1}{(k_{0}+i\epsilon)(k_{0}+i\epsilon)-\vec{k}^{2}-m^{2}}
\end{equation}
still satisfies the causality condition, though this choice 
does not ensure positive energy condition nor positive norm
condition. This fact shows that the causality is a condition
independent from other two basic conditions. Incidentally, 
 $T^{\star}$-product and 
$T$-product do not always agree with each other. For example, if 
one considers 
\begin{eqnarray}
{\cal L}=-im^{2}\xi(x)\eta(x)
\end{eqnarray}
the path integral gives
\begin{eqnarray}
\int d^{4}x\exp[ik(x-y)]\langle0|T^{\star}\hat{\xi}(x)
\hat{\eta}(y)|0\rangle
=\frac{1}{m^{2}-i\epsilon}
\end{eqnarray}
and BJL prescription gives 
\begin{eqnarray}
\int d^{4}xe^{ik(x-y)}\langle0|T\hat{\xi}(x)
\hat{\eta}(y)|0\rangle
&=&\int d^{4}xe^{ik(x-y)}\langle0|T^{\star}\hat{\xi}(x)
\hat{\eta}(y)|0\rangle\\
&-&\lim_{k^{0}\rightarrow\infty}\int d^{4}x
e^{ik(x-y)}\langle0|T^{\star}\hat{\xi}(x)\hat{\eta}(y)|0\rangle
=0.\nonumber
\end{eqnarray}
We also note that if one imposes the 
positive norm condition for abnormal spin-statistics assignment 
in operator formalism\cite{bjorken2}, one can of course detect 
the acausal behavior of the time ordered product by BJL 
prescription.

If one integrates over the variable $k^{0}$ in (3.13), 
one can write the propagator as 
\begin{eqnarray}
&&\langle0|T\hat{\xi}(x)\hat{\eta}(y)|0\rangle\nonumber\\
&&=\theta(x^{0}-y^{0})\langle0|\hat{\xi}(x)\hat{\eta}(y)|0\rangle
-\theta(y^{0}-x^{0})\langle0|\hat{\eta}(y)\hat{\xi}(x)|0\rangle
\nonumber\\
&&=(i)
\int\frac{d^{4}k}{(2\pi)^{4}}\exp[-ik(x-y)]
\frac{i}{k_{\mu}k^{\mu}-m^{2}+i\epsilon}\nonumber\\
&&=(i)\{\theta(x^{0}-y^{0})
\int\frac{d^{3}k}{(2\pi)^{3}2\omega}\exp[-i\omega(x^{0}-y^{0})+
i\vec{k}(\vec{x}-\vec{y})]\nonumber\\
&&\ \ \ +\theta(y^{0}-x^{0})
\int\frac{d^{3}k}{(2\pi)^{3}2\omega}\exp[-i\omega(y^{0}-x^{0})+
i\vec{k}(\vec{y}-\vec{x})]\}
\end{eqnarray}
with $\omega=\sqrt{\vec{k}^{2}+m^{2}}$. Note that both of the 
terms with $\theta(x^{0}-y^{0})\exp[-i\omega(x^{0}-y^{0})]$ and
 $\theta(y^{0}-x^{0})\exp[-i\omega(y^{0}-x^{0})]$ ensure
the positive energy condition, which is a result of Feynman's
$m-i\epsilon$ prescription. 
The presence of the extra imaginary factor $i$ in the right-hand 
side of (3.26)
thus shows  the indefinite inner product of $\xi$ and $\eta$ 
fields in the Hilbert space.  We note that this
$m^{2}-i\epsilon$ prescription and the propagator (3.13),
when applied to the ghosts in  gauge theory, are 
consistent with the BRST cohomology. See, for example,
Ref.\cite{fujikawa3}. 

If one uses the complex Grassmann variable 
$\varphi(x)=(\xi(x)+i\eta(x))/\sqrt{2}$ instead, (3.26) is 
replaced by 
\begin{eqnarray}
&&\langle0|T\hat{\varphi}(x)\hat{\varphi}^{\dagger}(y)
|0\rangle\nonumber\\
&&=\theta(x^{0}-y^{0})\langle0|\hat{\varphi}(x)
\hat{\varphi}^{\dagger}(y)|0\rangle
-\theta(y^{0}-x^{0})\langle0|\hat{\varphi}^{\dagger}(y)
\hat{\varphi}(x)|0\rangle
\nonumber\\
&&=(\frac{-i}{2})[\langle0|T\hat{\xi}(x)\hat{\eta}(y)|0\rangle
+\langle0|T\hat{\xi}(y)\hat{\eta}(x)|0\rangle]\nonumber\\
&&=
\int\frac{d^{4}k}{(2\pi)^{4}}\exp[-ik(x-y)]
\frac{i}{k_{\mu}k^{\mu}-m^{2}+i\epsilon}\nonumber\\
&&=\theta(x^{0}-y^{0})
\int\frac{d^{3}k}{(2\pi)^{3}2\omega}\exp[-i\omega(x^{0}-y^{0})+
i\vec{k}(\vec{x}-\vec{y})]\nonumber\\
&&\ \ \ +\theta(y^{0}-x^{0})
\int\frac{d^{3}k}{(2\pi)^{3}2\omega}\exp[-i\omega(y^{0}-x^{0})+
i\vec{k}(\vec{y}-\vec{x})]
\end{eqnarray}
and the negative norm in the sense of operator formalism appears 
in the second term of the time ordered product: Namely, if one 
expands 
\begin{equation}
\hat{\varphi}(x)=\int\frac{d^{3}k}{\sqrt{(2\pi)^{3}2\omega}}[
\hat{a}_{k}e^{-ikx}+\hat{b}^{\dagger}_{k}e^{ikx}]
\end{equation}
one obtains 
$\langle0|\hat{b}_{k}\hat{b}^{\dagger}_{k}|0\rangle<0$.
 (If one uses
the ordinary complex numbers for $\hat{\varphi}(x)$, the 
right-hand side of (3.27) remains the same but the coefficient
of the term $\theta(y^{0}-x^{0})\langle0|
\hat{\varphi}^{\dagger}(y)\hat{\varphi}(x)|0\rangle$ changes sign
and the positive norm condition is satisfied. The causality is
also satisfied for a general choice of $i\epsilon$ prescription.)
What this means is that the operator transcription induced by 
path integration
\begin{eqnarray}
\varphi(x)\rightarrow\hat{\varphi}(x),\nonumber\\
\varphi^{\dagger}(x)\rightarrow \hat{\varphi}^{\dagger}(x)
\end{eqnarray}
ensures  the hermitian conjugation 
$\hat{\varphi}^{\dagger}(x)=
(\hat{\varphi}(x))^{\dagger}$ in the operator sense for the
complex numbers but not for the Grassmann numbers. 

An attempt to quantize a  spin-$0$ field by using
 Grassmann variables thus either leads to a vanishing action or 
an indefinite metric in Hilbert space. 

In passing, we comment on the quantization of Maxwell field (and
also general Yang-Mills fields).
In the Feynman gauge, one deals with the Lagrangian
\begin{equation}
{\cal L}=-\frac{1}{2}\partial_{\alpha}A_{\mu}\partial^{\alpha}
A^{\mu}
\end{equation}
and thus the use of Grassmann variables for $A_{\mu}$ leads to
a trivial theory. Besides, a consistent description of all the 
classical electromagnetic phenomena is lost in such a case.
The Maxwell field need to be quantized in terms of ordinary
real numbers.

\section{Statistics for Dirac particles}

To analyze the statistics for Dirac particles, we examine
the QED-type Lagrangian
\begin{equation}
{\cal L}=\bar{\psi}i\Dslash\psi-m\bar{\psi}\psi
\end{equation}
where
\begin{equation}
\Dslash=\gamma^{\mu}(\partial_{\mu}-ie_{0}A_{\mu}).
\end{equation}
Our metric convention is $g_{\mu\nu}=(1,-1,-1,-1)$, and the 
$4\times 4$ $\gamma^{\mu}$ matrices satisfy the relation
$(\gamma^{\mu})^{\dagger}=\gamma_{\mu}$, namely, the spatial
components $\gamma^{k}$ are anti-hermitian.

The path integral is defined by
\begin{equation}
\int{\cal D}\bar{\psi}{\cal D}\psi\exp\{i\int d^{4}x
[\bar{\psi}i\Dslash\psi-m\bar{\psi}\psi]\}.
\end{equation}
As for the propagator, we have
\begin{equation}
\langle 0|T^{\star}\hat{\psi}(x)\hat{\bar{\psi}}(y)|0\rangle
=\frac{i}{i\Dslash-m+i\epsilon}\delta^{4}(x-y)
\end{equation}
which does not distinguish the Grassmann or ordinary complex
numbers for the field variables. 
Here we used Feynman's $m-i\epsilon$ prescription, which ensures a
smooth continuation from Euclidean theory. 

The derivation
of this propagator is important and we give a detailed account.
We start with the path integral with source terms added
\begin{equation}
Z(\eta, \bar{\eta})=\int{\cal D}\bar{\psi}{\cal D}\psi
\exp\{i\int d^{4}x{\cal L}_{\eta}\}
\end{equation} 
with
\begin{equation}
{\cal L}_{\eta}=\bar{\psi}i\Dslash\psi-m\bar{\psi}\psi
+\bar{\eta}\psi+\bar{\psi}\eta.
\end{equation}
We make the following change of variables in this 
${\cal L}_{\eta}$
\begin{eqnarray}
&&\psi(x)\rightarrow \psi^{\prime}(x)
-\frac{1}{i\Dslash-m+i\epsilon}\eta,
\nonumber\\
&&\bar{\psi}(x)\rightarrow \bar{\psi}^{\prime}(x)-
\bar{\eta}\frac{1}{i\Dslash-m+i\epsilon}
\end{eqnarray}
and we obtain
\begin{equation}
{\cal L}_{\eta}=\bar{\psi}^{\prime}[i\Dslash-m]\psi^{\prime}
-\bar{\eta}\frac{1}{i\Dslash-m+i\epsilon}\eta.
\end{equation}
We thus have
\begin{eqnarray}
Z(\eta, \bar{\eta})&=&\int{\cal D}\bar{\psi}{\cal D}\psi
\exp\{i\int d^{4}x[\bar{\psi}^{\prime}[i\Dslash-m]\psi^{\prime}
-\bar{\eta}\frac{1}{i\Dslash-m+i\epsilon}\eta]\}\nonumber\\
&=&\int{\cal D}\bar{\psi}^{\prime}{\cal D}\psi^{\prime}
\exp\{i\int d^{4}x[\bar{\psi}^{\prime}[i\Dslash-m]\psi^{\prime}
-\bar{\eta}\frac{1}{i\Dslash-m+i\epsilon}\eta]\}
\end{eqnarray}
where we used the translational invariance of the measure
${\cal D}\bar{\psi}{\cal D}\psi
={\cal D}\bar{\psi}^{\prime}{\cal D}\psi^{\prime}$.
The propagator is then calculated for the Grassmann variables as 
(for which $\eta$ and $\bar{\eta}$ are also Grassmann numbers)
\begin{eqnarray}
\langle 0|T^{\star}\hat{\psi}(x)\hat{\bar{\psi}}(y)|0\rangle&=&
\frac{1}{Z}\int{\cal D}\bar{\psi}{\cal D}\psi 
\psi(x)\bar{\psi}(y)\exp\{i\int d^{4}x
[\bar{\psi}i\Dslash\psi-m\bar{\psi}\psi]\}\nonumber\\
&=&\frac{1}{i}\frac{\delta}{\delta \bar{\eta}(x)}
\frac{-1}{i}\frac{\delta}{\delta \eta(y)}\ln Z|_{\eta=0}
\nonumber\\
&=&\frac{i}{i\Dslash-m+i\epsilon}\delta^{4}(x-y)
\end{eqnarray}
where the operator expression in the left-hand side follows
from Schwinger's action principle.
Similarly, for complex numbered variables (for which $\eta$ and 
$\bar{\eta}$ are also complex numbers) we have
\begin{eqnarray}
\langle 0|T^{\star}\hat{\psi}(x)\hat{\bar{\psi}}(y)|0\rangle&=&
\frac{1}{i}\frac{\delta}{\delta \bar{\eta}(x)}
\frac{1}{i}\frac{\delta}{\delta \eta(y)}\ln Z|_{\eta=0}
\nonumber\\
&=&\frac{i}{i\Dslash-m+i\epsilon}\delta^{4}(x-y).
\end{eqnarray}
These propagators exhibit precisely the local structure of the 
Lagrangian, since the path integral measure is local.
The propagator specifies the norm of single particle states, and
Feynman's $m-i\epsilon$ prescription dictates the propagation of
 negative energy states in the negative time direction and 
thus ensures the positive energy condition.
These relations, (4.10) and (4.11), show  that we cannot 
identify the statistics for Dirac particles 
by just looking at the right-hand side of propagators only.

There are thus two alternative paths to analyze the 
spin-statistics theorem in  path integral formalism.
The first path, which corresponds to the path taken by 
Pauli\cite{pauli2}, is to combine Schwinger's action principle
with path integral. The second path taken by 
Feynman\cite{feynman} will be discussed later. 

We first establish the causality by 
considering the Fourier transform of the above Feynman propagator
 defined by time ordered product. By using 
$\delta^{4}(x-y)=\int\frac{d^{4}k}{(2\pi)^{4}}
\exp[-ik(x-y)]$  we obtain the identical expression 
for these two cases in the right-hand side for a {\em free field} 
propagator (with spinor indices explicitly written)
\begin{equation}
\langle 0|T^{\star}\hat{\psi}_{\alpha}(x)
\hat{\bar{\psi}}_{\beta}(y)|0\rangle
=\int\frac{d^{4}k}{(2\pi)^{4}}\exp[-ik(x-y)]
(\frac{i}{\kslash-m+i\epsilon})_{\alpha\beta}.
\end{equation}
By employing BJL prescription as we did for a scalar particle, we
find in the present case as 
\begin{eqnarray}
&&\delta(x^{0}-y^{0})\langle 0|[\hat{\psi}_{\alpha}(x),
\hat{\bar{\psi}}_{\beta}(y)]_{\pm}|0\rangle
=\gamma^{0}_{\alpha\beta}\delta^{4}(x-y),\\
&&\delta(x^{0}-y^{0})\langle 0|[\hat{\psi}_{\alpha}(x),
\hat{\psi}_{\beta}(y)]_{\pm}|0\rangle=
\delta(x^{0}-y^{0})\langle 0|[\hat{\bar{\psi}}_{\alpha}(x),
\hat{\bar{\psi}}_{\beta}(y)]_{\pm}|0\rangle=0\nonumber
\end{eqnarray}
where $\pm$ correspond to the Grassmann or complex
numbers, respectively. 
The first relation in (4.13) is obtained from 
\begin{eqnarray}
&&(-ik^{0})\int d^{4}x\exp[ik(x-y)]
\langle 0|T\hat{\psi}_{\alpha}(x)
\hat{\bar{\psi}}_{\beta}(y)|0\rangle\nonumber\\
&&=
\int d^{4}x\exp[ik(x-y)]\nonumber\\
&&\times\frac{\partial}{\partial x^{0}}
[\theta(x^{0}-y^{0})\langle0|\hat{\psi}_{\alpha}(x)
\hat{\bar{\psi}}_{\beta}(y)|0\rangle
\mp\theta(y^{0}-x^{0})\langle0|
\hat{\bar{\psi}}_{\beta}(y)\hat{\psi}_{\alpha}(x)|0\rangle]
\nonumber\\
&&=\int d^{4}x\exp[ik(x-y)]\delta(x^{0}-y^{0})
\langle0|[\hat{\psi}_{\alpha}(x),
\hat{\bar{\psi}}_{\beta}(y)]_{\pm}|0\rangle\nonumber\\
&&+
\int d^{4}x\exp[ik(x-y)]\langle0|T\partial_{x^{0}}
\hat{\psi}_{\alpha}(x)
\hat{\bar{\psi}}_{\beta}(y)|0\rangle\nonumber\\
&&= k^{0}(\frac{1}{\kslash-m+i\epsilon})_{\alpha\beta}
\end{eqnarray}
and considering the limit $k^{0}\rightarrow\infty$. 
The second relations in (4.13) arise since 
there are no propagators for , for example, 
$\langle 0|T^{\star}\hat{\psi}_{\alpha}(x)
\hat{\psi}_{\beta}(y)|0\rangle=0$. These expressions (4.13) show 
that 
the causality (Condition 2) is always ensured for either choice
of variables. We note that the causality condition is satisfied
for other choices of $i\epsilon$ prescription, though such 
choices violate the positive energy condition: See (3.22). 

We now examine the norm in Hilbert space by performing integral 
over $k^{0}$ in the propagator  as
\begin{eqnarray}
&&\langle 0|T\hat{\psi}_{\alpha}(x)
\hat{\bar{\psi}}_{\beta}(y)
|0\rangle\\
&&=\theta(x^{0}-y^{0})
\int\frac{d^{3}k}{(2\pi)^{3}2\omega}e^{-i\omega(x^{0}-y^{0})+
i\vec{k}(\vec{x}-\vec{y})}\sum_{s} u_{\alpha}(k,s)
\bar{u}_{\beta}(k,s)
\nonumber\\
&&\ \ -\theta(y^{0}-x^{0})
\int\frac{d^{3}k}{(2\pi)^{3}2\omega}e^{-i\omega(y^{0}-x^{0})+
i\vec{k}(\vec{y}-\vec{x})}\sum_{s} v_{\alpha}(k,s)
\bar{v}_{\beta}(k,s)\nonumber
\end{eqnarray}
where we defined $k^{0}$ anew as $k^{0}=\omega=\sqrt{\vec{k}^{2}
+m^{2}}$. Note that both of the 
terms with $\theta(x^{0}-y^{0})\exp[-i\omega(x^{0}-y^{0})]$ and
 $\theta(y^{0}-x^{0})\exp[-i\omega(y^{0}-x^{0})]$ ensure
the positive energy condition, which is a result of Feynman's
$m-i\epsilon$ prescription. 
We normalize the positive energy $u_{\alpha}(k,s)$ and
negative energy $v_{\alpha}(k,s)$ spinor solutions as
\begin{eqnarray}
&&\sum_{s} u_{\alpha}(k,s)\bar{u}_{\beta}(k,s)
=(\kslash+m)_{\alpha\beta},\nonumber
\\
&&\sum_{s} v_{\alpha}(k,s)\bar{v}_{\beta}(k,s)
=(\kslash-m)_{\alpha\beta}.
\end{eqnarray}
On the other hand, in conjunction with Schwinger's action
principle,  we have a
different time ordering property for the Grassmann and ordinary
numbers in the left-hand side:
\begin{eqnarray}
&&\langle0|T\hat{\psi}_{\alpha}(x)
\hat{\bar{\psi}}_{\beta}(y)|0\rangle\\
&&=\theta(x^{0}-y^{0})\langle0|\hat{\psi}_{\alpha}(x)
\hat{\bar{\psi}}_{\beta}(y)|0\rangle\mp 
\theta(y^{0}-x^{0})\langle0|\hat{\bar{\psi}}_{\beta}(y)
\hat{\psi}_{\alpha}(x)|0\rangle\nonumber
\end{eqnarray}
where the first minus sign corresponds to the Grassmann number 
and the second plus sign corresponds to the ordinary complex 
number.
We know that the Grassmann choice gives rise to the positive
normed inner product for both of the electron and
the positron. This is understood by considering  
$\langle 0|T\hat{\psi}_{\alpha}(x)
\hat{\psi}^{\dagger}_{\alpha}(y)
|0\rangle $ instead of 
$\langle 0|T\hat{\psi}_{\alpha}(x)
\hat{\bar{\psi}}_{\beta}(y)
|0\rangle$ in (4.15) and (4.17). The Fourier transform of the 
right-hand side 
is identical for both choices, but the second term in the time 
ordering changes sign for the ordinary complex number. This 
means that the second term in the time ordering, which 
corresponds to the positron, acquires negative norm in the sense 
of conventional operator formalism\cite{pauli2} if one uses 
complex numbers: Namely, if one 
expands 
\begin{equation}
\hat{\psi}_{\alpha}(x)=\sum_{s}\int
\frac{d^{3}k}{\sqrt{(2\pi)^{3}2\omega}}
[\hat{a}_{k}(s)e^{-ikx}u_{\alpha}(k,s)
+\hat{b}^{\dagger}_{k}(s)e^{ikx}v_{\alpha}(k,s)]
\end{equation}
one obtains 
$\langle0|\hat{b}_{k}(s)\hat{b}^{\dagger}_{k}(s)|0\rangle<0$. 

What this means is that the operator transcription induced by 
path integration
\begin{eqnarray}
\psi(x)\rightarrow\hat{\psi}(x),\nonumber\\
\psi^{\dagger}(x)\rightarrow
\hat{\psi}^{\dagger}(x)
\end{eqnarray}
ensures the hermitian conjugation 
$\hat{\psi}^{\dagger}(x)=
(\hat{\psi}(x))^{\dagger}$ in the operator sense for the
Grassmann numbers but not for the complex numbers. 
The use of ordinary complex numbers for 
Dirac particles thus violates the positive norm condition 
(Condition 3).

\section{Euclidean analysis}

Feynman in his analysis of the spin-statistics 
theorem\cite{feynman} took a path, which is more in line 
with the spirit of path integral formalism,
and looked for other inconsistencies if one applies the abnormal
statistics.  The Feynman's $i\epsilon$ 
prescription allows a smooth continuation to Euclidean theory
by Wick rotation $x^{0}\rightarrow -ix^{4}$ with real $x^{4}$, 
and the actual calculations in path integral are usually 
performed in Euclidean setting.
The Euclidean theory thus obtained is regarded to satisfy the 
positive energy condition, since we can at any time rotate the 
metric back to Minkowski one up to any finite order in 
perturbation. 
Similarly, the time ordered product in Minkowski theory is 
readily recovered from Euclidean theory and thus the causality
is regarded to be ensured. The basic criterion is thus the 
positive norm condition.
 
We now follow the path of Feynman, though 
our analysis is not quite identical to that of Feynman, and look
for other inconsistencies if one applies the abnormal 
statistics in Euclidean theory 
\begin{equation}
\int{\cal D}\bar{\psi}{\cal D}\psi[{\cal D}A_{\mu}]
\exp\{\int d^{4}x[\bar{\psi}i\Dslash\psi-m\bar{\psi}\psi
-\frac{1}{4}F_{\mu\nu}F^{\mu\nu}]\}.
\end{equation}
Here a suitable gauge fixing is included in $[{\cal D}A_{\mu}]$.
As a phenomenon which is 
sensitive to the choice of path integral variables, we illustrate
 the use of the Weyl anomaly. The Weyl anomaly is basically a
one (or higher)-loop effect, but it is treated as if it were a 
tree-level effect in path integral\cite{fujikawa}. 
The Weyl transformation is defined by $g_{\mu\nu}(x)\rightarrow
\exp[-2\alpha(x)]g_{\mu\nu}(x)$ and the Dirac fields are 
transformed as 
\begin{eqnarray}
&&\bar{\psi}(x)\rightarrow\bar{\psi}^{\prime}(x)=
\exp[-\frac{1}{2}\alpha(x)]\bar{\psi}(x),
\nonumber\\
&&\psi(x)\rightarrow\psi^{\prime}(x)=\exp[-\frac{1}{2}\alpha(x)]
\psi(x).
\end{eqnarray}
This is the Weyl transformation for weight $1/2$ 
variables $\tilde{\psi}(x)=(g)^{1/4}\psi(x)$  and 
$\tilde{\bar{\psi}}(x)=(g)^{1/4}\bar{\psi}(x)$
in the Euclidean flat space-time limit, which differs from the 
naive Weyl transformation, for example, 
$\psi(x)\rightarrow\psi^{\prime}(x)=\exp[\frac{3}{2}\alpha(x)]
\psi(x)$.
We then obtain  a Jacobian factor (for the 
general coordinate invariant measure 
${\cal D}\tilde{\bar{\psi}}{\cal D}\tilde{\psi}$ in the Euclidean
flat space-time limit)\cite{fujikawa}
\begin{equation}
{\cal D}\bar{\psi}^{\prime}{\cal D}\psi^{\prime}
=J(\alpha) {\cal D}\bar{\psi}{\cal D}\psi
\end{equation}
where
\begin{equation}
J(\alpha)=\exp[\pm\int d^{4}x\alpha(x)\frac{e_{0}^{2}}{24\pi^{2}}
F^{\mu\nu}F_{\mu\nu}].
\end{equation}
The coefficients of this Weyl anomaly, in particular,
$\pm$ signs correspond to the Grassmann number or the ordinary
complex number, respectively. 
The sign difference appears from the fundamental property of 
the path integral measure, namely, the measure is defined by 
(left-)derivatives for Grassmann variables.  

The coefficient of the Weyl anomaly
is thus a good indicator of the statistics of particles.
The coefficient of the Weyl (or trace) anomaly, which is related 
to scale transformation\cite{coleman}, gives the (lowest order) 
$\beta$ function of the renormalization group\cite{adler},
$\beta(e)=\pm\frac{e^{3}}{12\pi^{2}}$, or if one treats $e^{2}$
as the coupling constant
\begin{equation}
\beta(e^{2})=\pm\frac{e^{4}}{6\pi^{2}}.
\end{equation}
The Grassmann number 
gives rise to the positive signature and asymptotically non-free 
theory, and the complex
number gives rise to the negative signature and asymptotically 
free theory.
It is known that the positive norm in the Hilbert space gives 
rise to asymptotically non-free theory for QED\cite{zee}:
A formal argument for this is based on the relation
\begin{equation}
e^{2}=Z^{-2}_{1}Z^{2}_{2}Z_{3}e^{2}_{0}
\end{equation}
where $e_{0}$ stands for the bare charge. The Ward identity
gives $Z_{1}=Z_{2}$, and the K\"{a}llen-Lehmann 
bound\cite{kallen} gives
$0\leq Z_{3}\leq 1$ for the photon wave function renormalization
factor, which is a result of the positive norm 
condition. We emphasize that the actual calculation of the wave
function renormalization factor is performed in Euclidean 
momentum space. 
We thus expect for small $e$
\begin{equation}
Z_{3}=1-ae_{0}^{2}\ln(\Lambda/m_{0})+....
\end{equation}
with a positive constant $a$, and we have
\begin{equation}
\beta(e^{2})=m_{0}\frac{\partial e^{2}}{\partial m_{0}}|_{\Lambda,
e^{2}_{0}}=ae^{4}>0
\end{equation}
to this order. 
The use of ordinary complex numbers for Dirac particles thus
contradicts the positive norm condition $0\leq Z_{3}\leq 1$.
This analysis is valid up to any finite order in perturbation 
theory for a sufficiently small coupling constant.

If one analyzes a complex scalar field theory defined by
\begin{equation}
{\cal L}=
[(\partial_{\mu}-ie_{0}A_{\mu})\varphi(x)]^{\dagger}
[(\partial^{\mu}-ie_{0}A^{\mu})\varphi(x)]
-m_{0}^{2}\varphi(x)^{\dagger}\varphi(x)
-\delta\lambda(\varphi(x)^{\dagger}\varphi(x))^{2}
\end{equation}
and the path integral
\begin{equation}
\int{\cal D}\varphi{\cal D}\varphi^{\dagger}[{\cal D}A_{\mu}]
\exp\{\int d^{4}x[{\cal L}-\frac{1}{4}F_{\mu\nu}F^{\mu\nu}]\},
\end{equation}
where the term with $\delta\lambda=O(e^{4}_{0})$ is a 
counter term to eliminate the induced
$(\varphi(x)^{\dagger}\varphi(x))^{2}$ coupling, 
the Weyl transformation of a weight $1/2$ variable 
$\tilde{\varphi}(x)=(g)^{1/4}\varphi(x)$ is given by
\begin{eqnarray}
&&\tilde{\varphi}(x)\rightarrow\tilde{\varphi}^{\prime}(x)=
\exp[-\alpha(x)]\tilde{\varphi}(x),
\nonumber\\
&&\tilde{\varphi}^{\dagger}(x)\rightarrow
(\tilde{\varphi}^{\dagger})^{\prime}(x)=\exp[-\alpha(x)]
\tilde{\varphi}^{\dagger}(x).
\end{eqnarray}
The Jacobian factor for the general coordinate invariant
measure in the Euclidean flat space-time limit is then 
calculated as a straightforward generalization of the 
calculation in Ref.\cite{fujikawa} as
\begin{equation}
{\cal D}{\varphi^{\dagger}}^{\prime}{\cal D}\varphi^{\prime}
=\exp[\pm\int d^{4}x\alpha(x)\frac{e_{0}^{2}}{96\pi^{2}}
F^{\mu\nu}F_{\mu\nu} ] {\cal D}{\varphi^{\dagger}}{\cal D}\varphi
\end{equation}
where $\pm$ signatures correspond to the complex number or 
the Grassmann number, respectively. If one combines this result
with an analysis of the $\beta$-function and the 
K\"{allen}-Lehmann bound in the lowest order in
$e_{0}^{2}$, one can exclude the 
Grassmann variables for scalar particles.

We can thus analyze the spin-statistical theorem in Euclidean 
theory without referring to the norm of on-shell states directly.
Even if one does not know the precise coefficient of the Weyl 
anomaly, one can readily recognize the change of the signature of
 the Weyl anomaly for abnormal statistics in path integral and 
thus the appearance of inconsistency. 
As for the analysis of the spin-statistics theorem in Euclidean 
theory, see also Schwinger\cite{schwinger}.

Alternatively if one uses a perturbative language for a Dirac
particle in Minkowski
space-time, which is close 
to the original analysis of Feynman\cite{feynman}, the vacuum 
polarization tensor changes sign if one uses the complex number. 
Consequently, its absorptive part which gives the decay 
probability of a virtual time-like photon
\begin{equation}
\gamma\rightarrow e\bar{e}
\end{equation}
changes sign, and we obtain a negative decay probability.
For Feynman, the spin-statistics theorem meant a clear 
understanding of the origin of this minus sign\cite{duck}
\cite{feynman2}.
What we have shown above is that this change 
of sign has a root in the very definition of the 
measure in path integral formulation.

We next comment on the statistics of the scalar and Dirac 
particles in  two-dimensional (one space and one time) theory. 
The so-called bosonization \cite{coleman2} is well known in 
two-dimensional theory.
The 
bosonization does not imply that one can use a complex number
for a Dirac particle, for example. Rather it implies an 
equivalent description either by using a real scalar particle
or by using a Dirac particle in two-dimensional theory. In this 
respect,
we note that the (gravitational) Weyl anomaly induced by
a massless real scalar, which is described by a real number
$\phi(x)$, and a massless Dirac particle, which is described by 
 Grassmann numbers $\psi(x)$ and $\bar{\psi}(x)$, is identical.
The Weyl transformation of relevant path integral variables
in two-dimensional theory is defined by
\begin{equation}
\tilde{\phi}(x)=(g)^{1/4}\phi(x)\rightarrow
\tilde{\phi}^{\prime}(x)=\exp[-\alpha(x)]
\tilde{\phi}(x)
\end{equation}
and 
\begin{eqnarray}
&&\tilde{\bar{\psi}}(x)=(g)^{1/4}\bar{\psi}(x)
\rightarrow\tilde{\bar{\psi}}^{\prime}(x)=
\exp[-\frac{1}{2}\alpha(x)]\tilde{\bar{\psi}}(x),
\nonumber\\
&&\tilde{\psi}(x)=(g)^{1/4}\psi(x)\rightarrow
\tilde{\psi}^{\prime}(x)=\exp[-\frac{1}{2}\alpha(x)]
\tilde{\psi}(x).
\end{eqnarray}
The path integral measure for both cases (i.e., free particles in 
two-dimensional curved apace-time) changes under the Weyl 
transformation as\cite{polyakov}
\begin{equation}
d\mu\rightarrow d\mu^{\prime}=
\exp[-\int d^{2}x\alpha(x)\frac{1}{24\pi}\sqrt{g}R]d\mu
\end{equation}
where $d\mu={\cal D}\tilde{\phi}$ or $d\mu={\cal D}\tilde{\psi}
{\cal D}\tilde{\bar{\psi}}$.
Since gravitational field couples to all the matter fields
universally, this agreement of the Weyl anomaly, which specifies 
the central charge ($c=1$) of the Virasoro algebra,
 suggests (though
does not prove) the equivalence of a real scalar particle and a 
Dirac particle in two-dimensional theory. The abnormal statistics
 changes
the signature of the Weyl anomaly and thus the signature of the 
central charge,
which spoils the positive norm condition in the  representation 
of the Virasoro algebra\cite{ginsparg}. In this sense, one can 
exclude the abnormal  assignment of statistics for the scalar and
Dirac particles in two-dimensional 
theory also. 
\section{Discussion}

The local path integral measure and Lorentz invariant local
Lagrangian together with time ordered products ensure causality
regardless of statistics. The Feynman's $m-i\epsilon$ 
prescription ensures positive energy condition regardless of 
statistics. 
 We find the indefinite 
metric for spin $0$ particles if one uses  Grassmann variables 
 and the negative norm 
for  negative energy states if one uses complex numbers for Dirac 
particles. This is in accord with the operator analysis of 
Pauli\cite{pauli2}, in responce to the Feynman's analysis of 
the spin-statistics theorem\cite{feynman}.
In the framework of path integral proper, one need to go 
one more step further
to recognize the negative norm for the abnormal case. By this 
way, we naturally arrive at the original treatment of Feynman in 
his analysis of the spin-statistics theorem\cite{feynman}. We here
illustrated the use of the coefficient of Weyl anomaly as a 
characteristic indicator of statistics, which is a direct
consequence of the definition  of path integral measure and works
not only for 4-dimensional theory but also for 2-dimensional 
theory.
(As for the Lorentz invariance of path integral measure, one can
examine it precisely by analyzing the Jacobian for local Lorentz
transformation\cite{alvarez-gaume}.)

In passing, we note  that the bosonic Dirac particle
is practically used as the Pauli-Villars regulator for a 
fermionic Dirac 
particle in path integral formulation\cite{fujikawa2}, which 
cancels all the possible anomalous Jacobians. In the 
regularization of continuum path integral, the cancellation of 
the anomalous Jacobian factor is essential to justify  naive 
manipulations. 

As for the generality of our arguments, the analysis of free 
propagators is applicable to all the cases. The analysis of free
propagators, when combined with the notion of Feynman diagrams, 
is extended to any finite order in perturbation theory.  As for 
the 
analyses of Weyl anomaly or the positivity of scattering 
processes, we note that all the known 
elementary particles with spin $1/2$ couple to gauge fields in 
4-dimensional theory. 
One can thus choose a suitable $U(1)$ gauge field associated 
with a Cartan subalgebra to establish the spin-statistics 
theorem for these elementary spinors.

In conclusion, we have shown a simple and 
coherent proof of the spin-statistics theorem in the framework 
of modern path integral, which incorporates 
the Grassmann numbers as an integral part\cite{berezin}. The 
minus commutation
relation between one Bose and one Fermi field arises naturally  
in this framework, since the complex numbers and Grassmann 
numbers commute in the classical level. 
\\

I thank Y. Matsuo for a helpful comment on the Virasoro algebra.

\end{document}